\documentclass[twocolumn]{emulateapj}
\usepackage[hyperfootnotes=false,colorlinks=true,linkcolor=blue,citecolor=blue]{hyperref}

\slugcomment{ApJL, in press}

\shortauthors{Thommes, Nilsson \& Murray} \shorttitle{Giant planet formation}

\begin{document} 

\title{Overcoming migration during giant planet formation }
\author{Edward W. Thommes$^1$, Leif Nilsson$^2$ and Norman Murray$^1$}
\affil{$^1$Canadian Institute for Theoretical Astrophysics, University of Toronto, 60 St. George Street, Toronto, Ontario M5S 3H8, Canada}
\affil{$^2$Department of Physics, University of Toronto, 60 St. George Street, Toronto, Ontario M5S 1A7, Canada}
\email{thommes@cita.utoronto.ca}

\begin{abstract}
In the core accretion model, gas giant formation is a race between growth and migration; for a core to become a jovian planet, it must accrete its envelope before it spirals into the host star.  We use a multizone numerical model to extend 
our previous investigation of the ``window of opportunity" for gas giant formation within a disk.  When the collision cross-section enhancement due to core atmospheres is taken into account, we find that a broad range of protoplanetary disks posses such a window.   
\end{abstract}

\keywords{protoplanetary disks, origin: solar system --- planets:
formation, planets:extrasolar}

\section{Introduction}
\label{intro}
Core accretion and direct gravitational instability are the two main competing models for how gas giant planets form.  
The core accretion picture  appears to be favoured by both observations 
\citep{2005ApJ...622.1102F}
and theory 
\citep{2005ApJ...621L..69R}
but its viability in the face of putative rapid orbital decay of protoplanets by planet-disk interactions 
\citep{1997Icar..126..261W,2000MNRAS.315..823P,2002ApJ...565.1257T}
has long been a cause of worry.  
\citet[herafter TM06]{2006ApJ...644.1214T} 
showed that in fact, even with rapid migration operating, a window of opportunity for successful core accretion can open later in a protostellar disk's lifetime, once it has dissipated to the point that accretion and migration timescales become comparable.  Here, we build on this work, adopting a more sophisticated model which explicitly tracks the history of each individual protoplanet all through the era of core formation.  In this way, we are able to assess the role of protoplanet mergers (as opposed to just planetesimal accretion) in the formation of cores, as well as to make predictions about the actual number of potential cores a given system will form.  We go on to include the effect of gas atmospheres on the protoplanets, which increases their accretion cross-sections.  In agreement with previous results, we find this greatly speeds up accretion once protoplanet masses reach $\sim 10^{-1}$ Earth masses (M$_\oplus$).  As a result, core accretion is successful, notwithstanding type I migration, for a wide range of disk parameters.  In \S \ref{sec: model}, we describe the model.   In \S \ref{sec: results}, we perform a sample computation using the model.  In \S \ref{sec: enhanced}, we add core atmospheres to the model and then recompute the sample case.  In \S \ref{sec: disk properties}, we examine core formation likelihood as a function of disk properties.  Conclusions are presented in \S \ref{sec: end}.  

\section{The model}
\label{sec: model}
As in TM06, we adopt for the type I migration rate the result obtained by 
\cite{2002ApJ...565.1257T}.
Likewise, we again use a ``particle in a box" estimate of the protoplanet accretion rate, assuming it to be in the "oligarchic" regime of growth
\citep{1998Icar..131..171K,2000Icar..143...15K,2003Icar..161..431T}.
Also, we use the simplifying assumption of a uniform planetesimal mass $m$.  The accretion rate of a protoplanet is
\begin{equation}
\frac{dM}{dt} \approx 3 \frac{\Sigma_m}{2 H_m} \pi R_M^2 \left (1+\frac{v_{\rm esc}^2}{v_{\rm rel}^2} \right ) v_{\rm rel}
\end{equation}
where $\Sigma_m$ is the surface density of the planetesimal disk, $H_m$ is its scale height, $R_M$ is the protoplanet radius, $v_{\rm esc}$ is the escape velocity from the protoplanet's surface, and $v_{rel}$ is the average relative velocity between the protoplanet and a planetesimal.  
The planetesimal velocities are increased through gravitational stirring by the protoplanets, and reduced by aerodynamic drag from the gas disk; we take the corresponding rates from \cite{1993Icar..106..210I} and \cite{1976PThPh..56.1756A} respectively.  Unlike in TM06, we explicitly calculate the velocity evolution of the plantesimal disk, without assuming equlibrium values for the random velocities of the plantesimals.

We compute the evolution  of the plantesimal disk using a multizone approach, dividing it up into radial bins.  
Each timestep, we compute the accretion zone, as well as the stirred zone, of each protoplanet, noting the planetesimal disk annuli it encompasses.  The stirred zone has a radial width
\begin{equation} 
\Delta r_{\rm stir} \approx 4 r_M \left [ \frac{4}{3} \left (e^2_m + i^2_m \right )+ 12 \left ( \frac{r_M}{r_H} \right )^2\right ]^{1/2}
\label{eq: stirred zone width}
\end{equation}
\citep{1993Icar..106..210I}
where $r_M$ is the protoplanet's orbital radius, $r_H \equiv (M/3 M_*)^{1/3} r_M$ its Hill radius, and $M_*$ the mass of the central star.  The accretion zone has a width
$\Delta r_{\rm accr} \approx \Delta r_{\rm stir}/2$.
In the oligarchic regime of growth, adjacent protoplanets maintain a separation of $\sim 10 r_H$.  We start the protoplanets with this spacing, and merge a pair of neighbors whenever they come within $7 r_H$ of each other.  At early times, such mergers represent the fact that there are winners and losers as neighboring protoplanets compete for building material.  Caution is required in interpreting any mergers which occur at late times; we consider this issue in \S \ref{sec: results} below.

\section{Results}
\label{sec: results}
We begin by re-calculating the ``baseline" model of TM06 using the above multizone approach.  That case begins with a disk of mass $M_d=0.15 M_\odot$ around a Solar-mass star.  The disk has an $\alpha$-parameterized viscosity with $\alpha=10^{-2}$, an initial outer radius of 50 AU, and a scale height profile $H=0.047(r/{\rm 1 AU})^{5/4}$ AU.  The gas disk's time evolution is calculated analytically using the similarity solution of
\cite{1974MNRAS.168..603L}.
At $t=0$, the accompanying solids disk follows the surface density profile of the gas, with a metallicity [Fe/H]=0.25.  It consists of planetesimals with a uniform radius of 1 km.  We take the ``snow line" 
to be at 2.7 AU as in the model of 
\cite{1981PThPh..70...35H} (see TM06 for details).
As in TM06, we stop the calculation when the gas mass interior to 100 AU drops below one Jupiter mass ($t_{\rm 100 AU}=5.5$ Myrs), noting also the time at which the gas mass interior to 30 AU drops below this value ($t_{\rm 30 AU}=2.4$ Myrs).  As a new addition, we calculate the critical core mass $M_{\rm crit}$ for runaway gas accretion as in \cite{2000ApJ...537.1013I}, assuming an envelope grain opacity of $\kappa=0.4 \,{\rm  cm}^2{\rm  g}^{-1}$ (see \S \ref{sec: enhanced} below).  This expression for $M_{\rm crit}$ becomes zero when planetesimal accretion ceases, so we choose the mass for runaway gas accretion onset to be $M_{\rm rwy}=min(M_{\rm crit}, 10 M_\oplus)$; envelope growth time for a 10 M$_\oplus$ body not accreting planetesimals is only $\sim 10^5$ yrs \citep{2000ApJ...537.1013I}.  
\begin{figure}
\plotone{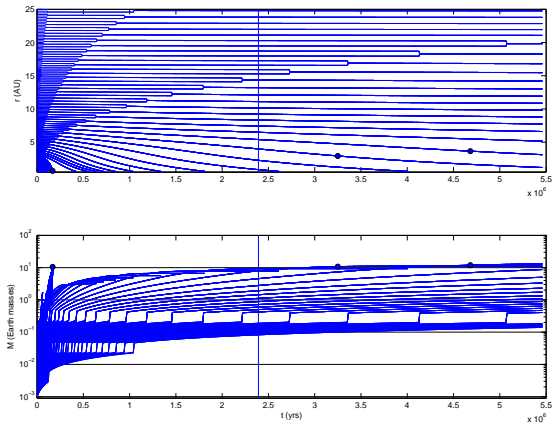}
\caption{Protoplanet migration (top) and growth (bottom) in an evolving protostellar disk with the parameters given in \S \ref{sec: results}.  The vertical line indicates $t_{\rm 30 AU}$, the time at which the gas disk mass in the inner 30 AU drops to 1 M$_{\rm Jup}$.  Filled circles show when a core first reaches $M_{\rm rwy}$, the runaway gas accretion mass.}
\label{fig: baseline_no_atmosphere}
\end{figure}

The result of the calculation is shown in Fig. \ref{fig: baseline_no_atmosphere}.  This can be directly compared to Fig. 6 of TM06.  The overall evolution matches our previous semianalytic result quite closely:  By $t_{\rm 30 AU}$, the largest protoplanet mass is just under 10 $M_\oplus$, and by $t_{\rm 100 AU}$ it is a bit over.  No protoplanets have reached $M_{\rm rwy}$ by $t_{\rm 30 AU}$, while two reach this threshold between $t_{\rm 30 AU}$ and $t_{\rm 100 AU}$, suggesting that core accretion growth of gas giants does have a chance to occur before the gas is lost.  
The multizone approach now allows us to see the role which mergers between neighbors plays in the growth of the protoplanets.  Both the $r$ vs. $t$ and $M$ vs. $t$ panels of Fig. \ref{fig: baseline_no_atmosphere} clearly show these mergers.  Because we start the protoplanets with a uniform separation in $r_H$, and adopt a uniform separation at which to merge neighbors, the mergers occur in successive, orderly generations which propagate outward through the disk over time.  From our chosen initial protoplanet mass of $10^{-3}$ M$_\oplus$, three generations of mergers occur, with the last one taking protoplanet masses from $\approx 0.2$ to $0.4$ M$_\oplus$.  Thus, protoplanet mergers play a significant role in growth (for a discussion, see 
\citealt{2006Icar..180..496C}),
 but only  up to a mass  $< 1$ M$_\oplus$ in the case shown here.
After that, the rate at which neighboring protoplanets diverge by type I migration exceeds the rate at which mass accretion expands their Hill radii, so that subsequent growth proceeds only by the sweep-up of the remaining planetesimals.  

An exception to the orderly sequences of mergers takes place within the first $2 \times 10^5$ years, beginning at the snow line.  Here, the wave of mergers starts to go inward with time rather than outward, with the result that one protoplanet accretes all its inner neighbors before falling off the inner edge of the disk.  It is able to do this because the density jump at the snow line is steep enough that within it, the growth timescale {\it decreases} outward.  This sets the scene for a ``predator" protoplanet to arise which grows faster than its inner neighbors.  Since the type I migration rate of a body is proportional to its mass, this body then overtakes its neighbors, accreting them all one by one, 
and finally exiting the inner edge of the disk just as it attains $M_{\rm rwy}$.
However, because the growth of the predator becomes dominated by mergers, and also because this is happening in a part of the disk where the majority of the planetesimal mass has already been swept up, we are no longer in the true oligarchic growth regime.  Thus our treatment of protoplanet mergers, being statistical in nature, loses its validity.  All we can say is that convergent migration of protoplanets takes place from the snow line inward.  Previous N-body work suggests that 
differential migration tends to result in the establishment of mean-motion resonances among protoplanets 
\citep{2005ApJ...626.1033T,2005AJ....130.2884M,TerqPap06Preprint}

\section{Atmosphere-enhanced accretion}
\label{sec: enhanced}

In the core-accretion model, a core becomes a gas giant when it undergoes runaway accretion from the surrounding gas disk, thus obtaining a massive atmosphere 
\citep{1978M&P....18....5C,1978PThPh..60..699M}.
Well before that time, however, the core will already start to accrete some gas 
\citep{1996Icar..124...62P}.
Planetesimals which intersect the atmosphere lose energy to aerodynamic drag; if they lose enough energy they are captured, even if their original trajectory would have caused them to miss the core
\citep{1996Icar..124...62P,2003A&A...410..711I,2003Icar..166...46I}.
Atmospheres thus provide an enhancement to the cores' planetesimal capture cross-sections, which increases with the density of the atmosphere and decreases with plantesimal size.  
As shown by the above authors, atmospheric capture can substantially boost a protoplanet's accretion rate beyond the early stages of its growth, once it is $\ga 10^{-1}$ M$_\oplus$.  

In order to add this effect to our calculation, we use the capture model of 
\cite{2006Icar..180..496C}
which is in turn a simplified version of the the model of 
\cite{2003A&A...410..711I}.
The atmosphere-enhanced capture radius $R_C$ is then given by
\begin{equation}
\left (\frac{R_C}{R_M} \right )^3 = \frac{0.0790 \mu^4 c R_M r_H}{\kappa r_m} \left ( \frac{M}{M_\odot}\right )^2 \left ( \frac{24}{24 + 5 \tilde{e}}\right )\left ( \frac{dM}{dt} \right )^{-1}
\label{eq: enhanced radius}
\end{equation}
where $\mu$ is the mean molecular weight of the gas (taken as 2.8, Solar composition), $c$ is the speed of light, $r_m$ is the planetesimal radius, $\kappa$ is the atmospheric opacity, and $\tilde{e} \equiv e r_M/r_H$.
We substitute $R_C$ for $R_M$ in our calculation, subject to the condition that $R_M < R_C < R_{\rm atmos}$, where $R_{\rm atmos}$ is the outer radius of the protoplanet's atmosphere, taken to be the smaller of its Hill and Bondi radii ($R_{\rm B} = G M/c_s^2$).
\begin{figure}
\plotone{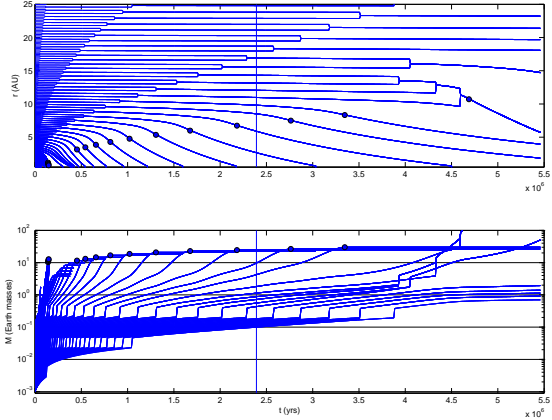}
\caption{Protoplanet migration (top) and accretion (bottom) for the same parameters as in Fig. \ref{fig: baseline_no_atmosphere}, but including the enhancement to the planetesimal capture cross-sections due to the growth of atmospheres on the protoplanets.}
\label{fig: baseline_with_atmosphere}
\end{figure}

We repeat the calculation of Fig. \ref{fig: baseline_no_atmosphere}, this time with atmosphere-enhanced capture radii for the protoplanets.  We use the opacity from the analytic fit of \cite{2003A&A...410..711I}, with a grain depletion factor of $f=0.1$ relative to the interstellar value to account for grain growth and settling \citep{2003Icar..165..428P}.  For simplicity, we use a single value for the opacity.  We conservatively choose the highest (i.e. lowest-temperature) value out of the range: $\kappa=4f$ cm $^{2}$g$^{-1}$ =0.4 cm$^{2}$g$^{-1}$.  
Fig. \ref{fig: baseline_with_atmosphere} shows the result.  Until protoplanet masses reach $\sim 10^{-1}$ M$_\oplus$, accretion proceeds in the same manner as before.  However, once the atmospheres become effective, growth quickly becomes much faster than in the unenhanced case.  Discounting the initial burst of growth at the edge of the snow line (see \S  \ref{sec: results} above), the first protoplanet to $M_{\rm rwy}$ does so at $4.5 \times 10^5$ yrs already.  By $t_{\rm 30 AU}=2.4$ Myrs, a total of eight protoplanets have managed to reach $M_{\rm rwy}$, each one as its planetesimal accretion ceases and $M_{\rm crit}$ decreases.  Since the later ones originate from further out in the disk, they survive for longer before migration removes them; the last four protoplanets reaching $M_{\rm rwy}$ before $t_{\rm 30 AU}$ last another $\sim 0.5-1$ Myrs.  Thus the window of opportunity for a core to become a gas giant is considerably larger than in the calculation without atmospheres, \S \ref{sec: results} above.  It should be noted that the calculation does not take into account various effects which become important at large core mass, in particular gap-opening and the extra mass contributed by the atmosphere, so any growth which happens at $M \gg 10$ M$_\oplus$ should not be taken too literally.  Likewise, the series of mergers starting inside 15 AU at $t \approx 4$ Myrs is subject to the caveat discussed in \S \ref{sec: results}. 

Calculations of concurrent core formation and migration were also performed by \cite{2005A&A...434..343A}, who concluded that type I migration would have to be reduced by an order of magnitude relative to the result of \cite{2002ApJ...565.1257T}  in order for core accretion to succeed.  The disparity between their results and ours appears to stem from two main sources:  First, they assumed a characteristic planetesimal size of 100 km rather than 1 km; this slows accretion relative to our calculation by reducing both the damping of planetesimal random velocities and atmosphere enhancement of protoplanet accretion radii.  It is worth noting that in a direct comparison using the same parameters, we actually obtain {\it slower} growth than \cite{2005A&A...434..343A}, likely because their code incorporates weaker  protoplanet-planetesimal gravitational stirring.  Secondly, they chose a rather large initial mass of 0.6 M$_\oplus$ for their protoplanets, with the result that at $t=0$, all of them are already migrating quite rapidly; throughout much of the disk, this removes the opportunity for disk dispersal to equalize growth and migration timescales.

\section{Giant planet formation as a function of disk properties}
\label{sec: disk properties}
\begin{figure}
\plotone{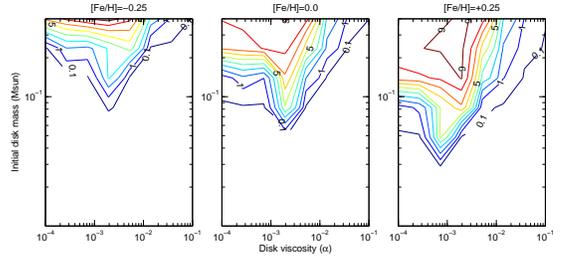}
\caption{The length of the time window $t_{\rm win}$ (contours, in Myrs) for the formation of gas giant planets as a function of initial disk mass $M_{d}$, $\alpha$ viscosity parameter, and metallicity [Fe/H].  The larger $t_{\rm win}$, the more likely it is that one or more gas giants will form in a given system; for $t_{\rm win} \la 10^5$ yrs, there is unlikely to be sufficient time.}
\label{fig: time_delay_contourplot_3panel}
\end{figure}

As in TM06, we now build up a global picture of how the likelihood of gas giant formation by core accretion changes as a function of disk properties, this time including the effect of core atmospheres.  We compute a grid of models, $0.01 $ M$_\odot \leq M_d \leq 0.4 $ M$_\odot$ and $10^{-4} \leq \alpha \leq 10^{-1}$, and repeat this for metallicities of [Fe/H]=-0.25, 0 and 0.25.  In TM06, the largest protoplanet mass at $t_{\rm 100 AU}$ was plotted for each model.  Here, we instead plot the length of the time window, $t_{\rm win}$, between the time the first protoplanet reaches $M_{\rm rwy}$, and $t_{\rm 30 AU}$; we consider this to be a better measure of how likely a given system is to succeed in forming a gas giant.  Also, we cut all computations off at 10 Myrs, even if $t_{\rm 30 AU} > 10$ Myrs, since this is the observational upper age limit for gas disks.  
Fig. \ref{fig: time_delay_contourplot_3panel} shows the results, indicating that $t_{\rm win}$ increases with [Fe/H], while the ``optimal" value of $\alpha$, which allows successful core accretion at the lowest disk mass, decreases with [Fe/H].  Within the range of observationally-inferred disk viscosities, $10^{-3} \la \alpha \la 10^{-2}$ 
\citep{1998ApJ...495..385H},
[Fe/H]=0.25 disks as low as $\sim 0.03$ M$_\odot$ in initial mass have $t_{\rm win} > 10^5$ yrs, and thus stand a chance of accreting a Jovian envelope in time 
\citep{2005Icar..179..415H}.
For disks with [Fe/H]=-0.25, $M_d \ga 0.08$ M$_\odot$ is required.  

\section{Conclusions}
\label{sec: end}

We have presented here a multizone calculation of concurrent planet growth and accretion in an evolving protoplanetary disk, focusing on whether giant planet cores can form in time.  The first part of our results agrees well with our previous semi-analytic model, while elucidating the role played by protoplanet-protoplanet mergers at early times. 
Since the calculation tracks individual protoplanets, we can now estimate the total number of bodies which reach core mass in a given system.  An important point is that although migration impedes planetesimal accretion by moving protoplanets to the already-depleted inner disk before they can finish their growth, this actually {\it helps} in attaining runaway gas accretion, provided a protoplanet is able to reach $\ga$ 10 M$_\oplus$ before its planetesimal supply is cut off.
 
When we add to our calculation a simple model for the effect of gas atmospheres on growing protoplanets, the results change dramatically.  In our baseline model, the time to grow a body large enough to serve as a giant planet core shrinks from over 3 million years to less than half a million years.  Thus in the competition between accretion and migration, the balance swings strongly 
toward the former, relatively speaking.   This makes the window of opportunity for the  formation of gas giants by core accretion---the time between the appearance of the first core-sized body, and the disappearance of the gas disk---much larger for a given set of parameters.  As a result, the range of disk properties which can plausibly produce gas giants, notwithstanding effective type I migration, becomes significantly broader than in the atmosphere-less calculation of TM06.  Now, we find that a Solar-metallicity disk of moderate initial mass ($M_d \ga 0.05$ M$_\odot$) stands a reasonable chance of giving birth to a jovian planet.  Since we neglect fragmentation of planetesimals, and the potentially very rapid accretion of fragments at low relative velocities dominated by Keplerian shear \citep{2004AJ....128.1348R}, which will also further increase the role of atmospheric capture 
\citep{2003A&A...410..711I},
the accretion rates in our calculation are likely conservative, and core formation may thus proceed even more readily; these effects will be added in future work.  

\acknowledgements{EWT thanks Shigeru Ida and Masahiro Ikoma for stimulating and informative discussions.  We also thank the referee for valuable comments, in particular for suggesting an improved treatment of critical core masses.}


\end{document}